\documentstyle[11pt,aaspp4,flushrt]{article}  
\def\be{\begin{equation}}
\def\ee{\end{equation}}

\def\msun{M_{\odot}}
\def\ergscc{\rm \  \ erg \ cm^{-3} \ s^{-1}}
\input psfig.tex
\begin{document}
\date{}
\title{Accretion flows: the role of the outer boundary condition}
\author{Feng Yuan}
\affil{Department of Astronomy, Nanjing University, Nanjing 210093,
China\\ Email: fyuan@nju.edu.cn }
\begin{abstract}
We investigate the influences of 
the outer boundary conditions(OBCs) 
on the structure of an optically thin accretion flow.
We find that OBC plays an important role in determining the topological
structure and the profiles of the density and temperature of the solution,
therefore it should be regarded as a new parameter in the accretion disk model.
\end{abstract}

 
\section{Introduction}
The behavior of an astrophysical accretion flow is described by a set 
of non-linear ordinary differential equations. 
Thus the outer boundary condition(OBC)
possibly plays an important role in such an {\it initial value problem}.
On the other hand, the complicated astrophysical environments
make the states of the accreting gas at the outer boundary, 
 such as its temperature and 
angular momentum, various. So it is necessary to investigate
the role of OBC in accretion process. 

However, this problem has not
received enough attention until now. In the standard thin disk 
model the differential equations are reduced to a set of linear algebraic
equations that don't entail any boundary conditions at all. 
In the later works on 
the {\it global} solutions for slim disks the thin disk solution
was usually adopted as the approximate OBC 
(Paczy\'nski \& Bisnovatyi-Kogan 1981; Muchoteb \& Paczy\'nski 1982;
 Matsumoto et al. 1984; Abramowicz et al. 1988; Chen \& Taam 1993).
In the case of optically thin accretion, this approach does not apply
any longer because of the importance of the energy advection throughout the
disk. In this case, self-similar solution by Narayan \& Yi(1994),
 or some ``arbitrarily'' adopted value around the self-similar
solution,
is usually chosen as the OBC(Narayan, Kato \& Honma 1997;
Chen, Abramowicz \& Lasota 1997; 
Nakamura et al. 1997; Manmoto, Mineshige \& Kusunose 
1997; Popham \& Gammie 1998).
They all concentrated on the solution under a specified boundary condition, 
while the role of OBC was not considered.
Shapiro(1973) is an exception.
In the context of spherically accretion,
he found that the temperature distribution and the radiation luminosity
of the disk depended sensitively on whether the interstellar medium
was an H I or an H II region.

In this {\it Letter}, taking the optically thin accretion
as an example, we investigate the effects of OBC
on the structure of the accreting flow.

\section{Equations and Results}
We consider a steady axisymmetric accretion flow around a black
hole of mass $M$($=10 \msun$ throughout this {\it Letter}),
employing a set of height-integrated
 equations widely used by the present workers. 
Paczy\'nski \& Wiita (1980) potential is used to mimic the
geometry of a Schwarzschild black hole. We first assume that the couple
between the electrons and ions is so strong that 
the flow is a one temperature plasma. The equations read as follows:
\be
-4\pi r H\rho v=\dot{M},
\ee
\be
v \frac{dv}{dr}=-\Omega_{\rm k}^2 r+\Omega^2 r-\frac{1}{\rho}\frac
{d}{dr}(\rho c_s^2),
\ee
\be
v(\Omega r^2-j)=\alpha c_s^2 r \frac{d{\rm ln}\Omega_k}{d{\rm ln}r},
\ee
\be
\frac{\rho v}{(\gamma-1)} \frac{dc_s^2}{dr}-c_s^2 v\frac{d\rho}
{dr}=q^+-q_{\rm br}^-.
\ee
All the quantities have their popular meanings.
The shear stress is taken to be simply proportional to the pressure, i.e., 
shear stress=$\alpha p\frac{d{\rm ln} \Omega_k}{d{\rm ln} r}$.
We assume the flow is optically thin, so 
the pressure is dominated by gas pressure,
$p=\frac{R}{\mu}\rho T$. 
For simplicity the bremsstrahlung emission $q_{\rm br}^-$
 is assumed to be the only 
cooling process. The viscous heating and radiation cooling rates are
\be
q^+=\alpha p \frac{d {\rm ln} \Omega_{\rm k}}{d {\rm ln}r}
\left(r\frac{d\Omega}{dr}\right),
\ee
\be
q^-_{\rm br}=6.2 \times 10^{20} \rho^2 T^{1/2} \ergscc,
\ee
respectively.

Since the accretion onto a black hole must be transonic, 
 the solution must satisfy the sonic point condition at 
a certain sonic radius $r_s$.
Besides this, we require 
the flow must satisfy two {\it outer boundary conditions} at a certain
outer boundary $r_{\rm out}$. (Note that the no torque condition required
at the black hole horizon is automatically 
satisfied under our viscosity description).
As $M, \dot{M}, \alpha$ are given, 
we obtain the global solution by adjusting the eigenvalue $j$ to ensure
that the flow satisfying the outer boundary condition
satisfies the sonic point condition as well. 
 
At $r_{\rm out}$, the two outer boundary conditions 
we imposed are $T=T_{\rm out}$ and 
$\lambda(\equiv \frac{v}{c_s} \equiv \frac{v}{\sqrt{p/\rho}}
)=\lambda_{\rm out}$. 
This set of boundary conditions are equivalent to $(T_{\rm out}, 
\Omega_{\rm out})$ according to eq. (3) 
because $j$ is not a free parameter. We assign
($T_{\rm out}, \lambda_{\rm out}$) to different sets of values 
and investigate their effects on the 
global solutions. The results are as follows.

 As $\dot{M}, \alpha$, and $r_{\rm out}$ are given, we find that 
only when $T_{\rm out}$ and $\lambda_{\rm out}$ 
are within a certain range do the global solutions exist.
For example, for the global solutions
with $\dot{M}=10^{-3}\dot{M}_{\rm E}$
$\alpha=10^{-2}$ and
$r_{\rm out}=10^3r_{\rm g}$ (Here the Eddington accretion 
rate reads $\dot{M}_{\rm E}=\frac{4 \pi GM}{c \kappa_{\rm es}}$
and $r_{\rm g}$ is the Schwarzschild radius), 
the values of $T_{\rm out}$ are close to the virial temperature
$T_{\rm virial}$($\equiv (\gamma-1)GMm/ \kappa r_{\rm out}$),
$T_{\rm out} \sim 0.1-1 T_{\rm virial}$. This range changes slightly
when we modify the values of $\dot{M}, \alpha$ and $r_{\rm out}$, 
and when we consider different radiation mechanisms.

 The solutions are remarkably different under different OBCs.
In terms of the values of $T_{\rm out}$, the whole parameter space 
spanned by $T_{\rm out}$ and $\lambda_{\rm out}$
can be divided into two regions.
The solutions located in the low-temperature region(referred to as type I)
 have the following
features, as the solid line in Fig. 1 shows: 1)
 the maximum feasible value of $\lambda_{\rm out}$ is close to 1; 
2) a maximum exists in the solution's angular momentum($l \equiv 
\Omega r^2$) profile;
3) the sonic radius $r_{\rm s}$ is always small;
and 4) the Bernoulli parameter
$Be$, which is equivalent to $\frac{1}{2}v^2+\frac{\gamma}{\gamma-1}c_s^2+\frac{l^2}{2 R^2}-\frac{1}
{R-2}$ is negative at large radii.
The solutions located in the high-temperature region can be 
further divided into two types in terms of the values of $\lambda_{\rm out}$. 
When $\lambda_{\rm out}$ is smaller
 than a critical value $\lambda_{\rm out,crit}$
(referred to as type II), 
the sonic radii of the corresponding
solutions are still small, but
the values of $Be$ become positive. Besides these, the angular momentum and 
the surface density($\Sigma \equiv 2 \rho H$)
 profiles and the topological structures(Mach number profile)
 differ greatly from those of type I, as shown by the dot-dashed line 
in Fig. 1.
 When $\lambda$ is greater than $\lambda_{\rm crit}$(referred to as type III),
 $r_{\rm s}$ suddenly becomes very large, as Table 1 and the dashed
line in Fig. 1 show.
This is a new kind of solution, since all previous works on ADAFs 
global solutions
only obtained solutions with small values of $r_{\rm s}$. 
This kind of solution is difficult to be found because
 the parameter space where 
it is located is too narrow.

It is interesting to note that such transition of the values of
$r_{\rm s}$ is similar to that in the adiabatic case(Abramowicz \& Zurek 1981).
In that case, $Be$ and $l$ are two constants of
motion. When $Be$ lies in a certain range, $r_{\rm s}$ will jump from
a small value to a large one when $l$ decreases across a
critical value $l_{\rm c}$, and the
 corresponding accretion patterns are
called disk-like(small $r_{\rm s}$, large $l$) 
and Bondi-like(large $r_{\rm s}$, small $l$) ones respectively. 
Our result indicates that such transition still exist 
although the flow becomes viscous(note that the increase of $\lambda$
corresponds to the decrease of the angular momentum). 

In general, 
 modifying $\lambda_{\rm out}$(i.e., modifying $\Omega_{\rm out}$)
 has much smaller effects on the solution than
modifying $T_{\rm out}$ does. The latter has a great effect on 
the topological structure and the profiles of the angular momentum
and surface density of the solutions. But its effect on the
temperature profile rapidly weakens with the decreasing radii, 
as the temperature profile plot in Fig. 1 shows.
We'll discuss this problem later.

The above calculation indicates that, besides $\dot{M}$ and $\alpha$,
OBC is also a crucial factor to determine 
the structure of the accretion flow. 

Present theoretical and observational works seem to support
the two temperature plasma assumption with the ions being much 
hotter than the electrons.
In the following we consider whether the effects of OBC
 are still important in this case. 
Comptonization of
the bremsstrahlung photon is also included in the cooling processes. 
The pressure and the energy equations should be replaced by,
\be
p=p_{\rm i} + p_{\rm e}= \frac{\rho k T_{\rm i}}{\mu_{\rm i}m_{\mu}}
+\frac{\rho k T_{\rm e}}{\mu_{\rm e}m_{\mu}},
\ee
\be
\rho v \left(\frac{d \varepsilon_i}{dr}+p_i d \left( \frac{1}{\rho}\right) 
\right)=q^+-q_{ie},
\ee
\be
\rho v \left(\frac{d \varepsilon_e}{dr}+p_e d \left( \frac{1}{\rho}\right) 
\right)=q_{ie}-q_{bc}^-,
\ee
where $\mu_{\rm i}$ and $\mu_{\rm e}$ denote the 
ion and electron molecular weights respectively,
$\varepsilon_i, \varepsilon_e$ are the internal energies of the ion
and electron per unit mass of the gas. Since for the temperature of interest,
the ion is always unrelativistic while the electron is transrelativistic,
following Shapiro(1973), we adopt the following:
\be
\varepsilon_i=\frac{3}{2}\frac{kT_i}{\mu_i m_{\mu}},
\hspace{1cm}  \varepsilon_{\rm e}=2\frac{kT_e}{\mu_e m_{\mu}}.
\ee
$q_{ie}$ in eq. (9) is the energy-transfer
 rate from the ions to electrons per unit volume
duo to Coulomb collisions(Dermer, Liang \& Canfield 1991):
\be
q_{ie}=\frac{3}{2}\frac{m_e}{m_i}n_en_i \sigma_T c {\rm ln} \Lambda 
\left(kT_i-kT_e\right)\frac{\left(\frac{2}{\pi}\right)^{1/2}+\left(\theta_e+
\theta_i\right)^{1/2}}{\left(\theta_e+\theta_i \right)^{3/2}},
\ee
where ${\rm ln} \Lambda=20$ is the Coulomb logarithm, and $\theta \equiv
kT/m c^2$ is the dimensionless temperature.
The thermal bremsstrahlung amplified by Comptonization $q_{bc}^-$ reads(
Rybicki \& Lightman 1979) as follows:
\be
q_{bc}^-=q_{\rm br}^- A,
\ee
\be
A={\rm max}\{\frac{3}{4}{\rm ln}^2[x_{ff}\sqrt{ln(x_{ff}^{-1})}],1\},
\ee
\be
x_{ff}=2.8\times 10^{12} T_e^{-7/4}\rho H^{1/2} .
\ee

Eqs. (7)-(14) together with eqs. (1)-(3) constitute the set of
equations describing a two temperature accretion flow.
In this case we must supply three physical quantities at $r_{\rm out}$,
 e.g. $T_{\rm out,i}, T_{\rm out,e}$ and $\lambda_{\rm out}
(\equiv v/\sqrt{p/ \rho})$.
Obviously, for a given $T_{\rm out,i}$, the upper limit of $T_{\rm out,e}$
is $T_{\rm out,i}$ while its lower limit is determined 
by $q_{\rm ie}=q_{\rm bc}^-$, i.e. all the energy transferred to the electrons
is radiated away.

We solve the above set of equations under different OBCs 
and find that the results are 
qualitatively the same, as shown by Fig. 2.
In this case we still find three types of solutions,
the effect of $\lambda_{\rm out}$ is still unimportant 
for an individual type of solution, and
the surface density profiles still differ greatly
for the solutions with different OBCs. Compared with the corresponding
solution in Fig. 1, the value of $Be$ of 
the type I solution (denoted by the solid line in Fig. 2)
decreases greatly. This  is
due to the inclusion of Comptonization 
together with the high surface density of the flow.

A remarkable new feature is that the electron temperature profiles
of the solutions with different OBCs(primarily $T_{\rm out,e}$) strongly 
differ from each other throughout the disk,
while for the ion 
 the discrepancies rapidly lessen with 
the decreasing radii from $r_{\rm out}$ like a 
one temperature plasma. 
This is because the electrons are essentially adiabatic in the 
present low $\dot{M}$ case, i.e., both $q_{\rm ie}$ and
$q_{\rm bc}^-$ are very small compared with the two terms on the left hand
side of eq. (9), so the electron temperature is {\it globally}
determined therefore the effect of OBC on $T_{\rm e}$
persists as the radius decreases.
But for the 
ion, as well as the one temperature plasma, the local
viscous dissipation 
in the energy equation plays an important role thus the energy equation
for ions is much more ``local'' than that for electrons, so 
the effect of OBC weakens rapidly with the decreasing radii.
Due to the same reason, we predict that OBC will be less important in 
determining the temperature profile of
an optically thick accretion flow, since both  the 
viscous dissipation and the radiation losses terms in the energy equation
are important thus the temperature will be {\it locally} determined.
But the flow should still present OBC-dependent dynamical behavior in 
, for example, the surface density and angular momenta profiles
 and the topological structure.

\section{Conclusions}
In this {\it Letter}, taking the optically thin one temperature
and two temperature plasma as examples, we investigate the role of
the outer boundary condition in the accretion process and find
that it plays an important role.
Our results indicate that in either case  
the whole solutions can be divided 
into three types in terms of their OBCs. 
 The topological structures and the profiles of
the angular momenta and the density are strongly 
different among the three types. As to the temperature we find that
for a two temperature plasma the discrepancies of $T_{\rm out,e}$
among the solutions with different OBCs will persist throughout the disk,
while for ions and a one temperature plasma 
the discrepancies of the temperature
rapidly lessen away from the outer boundary.
Such discrepancies among the dynamic quantities will
further induce the discrepancies of 
 the radiation luminosity and the emission spectra 
of the accretion disc. Therefore we argue
that OBC should be regarded as a new parameter in the accretion model. 

\acknowledgements

F.Y. thanks Dr. Kenji E. Nakamura for helpful discussions and
the anonymous referee for
useful suggestions to improve the presentation of this paper.
This work is supported in part by the National Natural Science
Foundation of China
under grants 19873005 and 19873007.

\references
\def\refpar{\hangindent=3em\hangafter=1}
\def\reference{\refpar\noindent}
\def\apj{ApJ}
\def\apjs{ApJS}
\def\mnras{MNRAS}
\def\aa{A\&A}
\def\aas{A\&A Suppl. Ser.}
\def\aj{AJ}
\def\araa{ARA\&A}
\def\nat{Nature}
\def\pasj{PASJ}

\reference  Abramowicz, M. A., Czerny, B., Lasota, J. P., \& Szuszkiewicz, E.,
1988, \apj, 332, 646

\reference Abramowicz, M., A., Zurek, W.H., 1981, ApJ, 246, 31

\reference Chen, X., Abramowicz, M. A., \& Lasota, J.-P. 1997, \apj, 476, 61
 
\reference Chen, X., \& Taam, R.E. 1993, \apj, 412, 254

\reference Dermer, C.D., Liang, E.P., \& Canfield, E. 1991, \apj, 369, 410

\reference Manmoto, T., Mineshige, S., Kusunose, M. 1997, \apj, 489, 791

\reference Matsumoto, R., Kato, S., Fukue, J., \& Okazaki, A.T. 1984, \pasj, 36,
71
 
\reference Muchotreb, B., \& Paczy\'nski, B. 1982, Acta Astron. 32, 1
 
\reference Nakamura, K. E., Kusunose, M., Matsumoto, R., \& Kato, S. 1997,
\pasj,  49, 503
 
\reference Narayan, R., Kato, S. \& Honma, F. 1997, \apj, 476, 49

\reference Narayan, R. \& Yi, I. 1994, \apj, 428, L13

\reference Paczy\'nski,B., \& Bisnovatyi-Kogan, G. 1981, Acta Astron., 31, 283

\reference Paczy\'nski, B., \& Wiita, P. J. 1980, \aa, 88, 23

\reference Popham, R., \& Gammie, C.F. 1998, \apj, 504, 419

\reference Rybicki, G., \& Lightman, A.P. 1979, Radiative Processes 
in Astrophysics (New York: Wiley)

\reference Shapiro, S.L. 1973, \apj, 180, 531
 
\newpage

\begin{figure}
\psfig{file=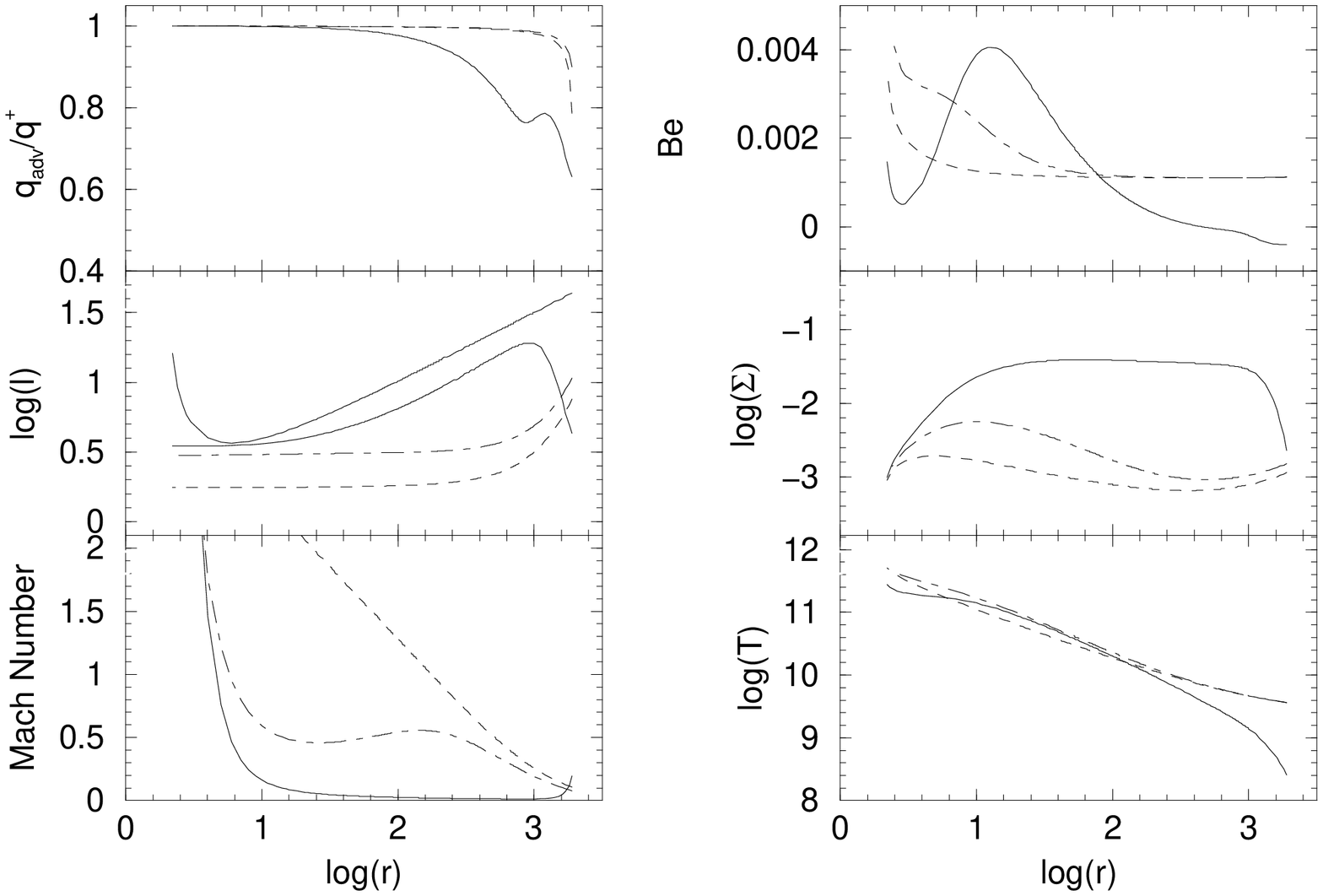,width=0.9\textwidth,angle=270}
\caption{
Solutions for one temperature global solutions with different OBCs for 
$\dot{M}=
10^{-3} \dot{M}_{\rm E}$ and $\alpha=10^{-2}$. The solid, dot-dashed and 
dashed lines represent $(T_{\rm out}, \lambda_{\rm out})$=($2 \times 
10^8 {\rm K}, 0.4), (3.6 \times 10^9 {\rm K}, 0.08)$ and 
$(3.6 \times 10^9 {\rm K}, 0.107$) respectively. The units of $\Sigma, T$
are ${\rm g \ cm^{-2}}$ and K, $r$, $Be$ and $l$ are in $c=G=M=1$ units.
Mach number is simply defined as $v/c_s$. The upper left-hand 
plot represents the ratio of the advected energy to the viscous
dissipated energy.}
\end{figure}

\begin{figure}
\psfig{file=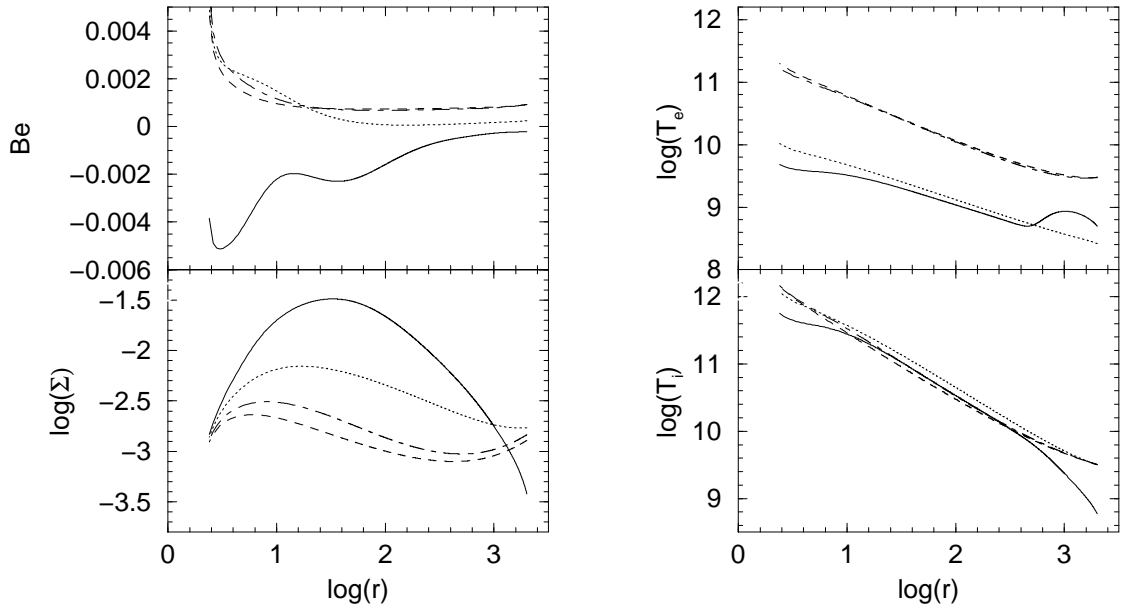,width=0.9\textwidth,angle=270}
\caption{
Solutions for two temperature accretion flows with different OBCs for 
$\dot{M}=
10^{-3}\dot{M}_{\rm E}$ and $ \alpha=10^{-2}$. The solid(type I),
dotted(type II), dashed(type III) and 
dot-dashed(type II) lines represent $(T_{\rm out,i},
 T_{\rm out,e}, \lambda)$=($6 \times 
10^8 {\rm K}, 5 \times 10^8 {\rm K}, 0.8), (3.2 \times 10^9 {\rm K},
2.62 \times 10^8 {\rm K}$ , $0.1),
(3.2 \times 10^9 {\rm K}, 3 \times 10^9 {\rm K}, 0.1)$ and $
(3.2 \times 10^9 {\rm K}, 3 \times 10^9 {\rm K}, 0.088)$, 
respectively. The units are same as those in Fig. 1. In the dotted line
$T_{\rm out,e}$ is determined
by setting $q_{\rm ie}=q_{bc}^-$ at $r_{\rm out}$.}
\end{figure}

\newpage
 
\begin{center}
Table 1.\\ Parameters, OBCs and the sonic radii of some solutions
\end{center}
 
\begin{center}
\begin{tabular}{cccclll} \hline \hline
$\dot{M}$ & $\alpha$ & $r_{\rm out}$ & $T_{\rm out}$ & $\lambda$ & $r_{\rm s}$
& $j$ \\
$(\dot{M}_{\rm E})$ & & $(r_{\rm g})$ & (K) & & $(r_{\rm g}/2)$ & $(r_{\rm g} c
/2)$\\
\hline
$10^{-3}$&$10^{-2}$&$10^3$&$2 \times 10^8$&0.2&4.5476&3.4962\\
&&&&0.4&4.5587&3.4887\\
&&&&0.8&4.5531&3.4841\\
&&&$3.6 \times 10^9$&0.08&5.7798&2.9869\\
&&&&0.1&6.2359&2.8973\\
&&&&0.102&155.1994&2.8023\\
&&&&0.107&176.3544&1.7578\\  \hline
$10^{-1}$&$10^{-1}$&$10^3$&$1.6 \times 10^9$&0.3&7.0615&2.3171\\
&&&&0.55&7.4851&2.2879\\
&&&&0.556&701.0070&2.1473\\
&&&&0.57&736.4865&1.2885\\   \hline
\end{tabular}
\end{center}

\end{document}